\documentclass[reqno,12pt,a4paper]{amsart}

\usepackage{mathrsfs}
\usepackage{amssymb}
\usepackage{amsfonts}
\usepackage{latexsym}
\usepackage{amsthm}
\usepackage[cp1251]{inputenc}
\usepackage{graphicx}
\def\lb{\label}

\newcommand{\er}[1]{\textrm{(\ref{#1})}}

\begin{document}


\renewcommand{\theequation}{\arabic{equation}}
\theoremstyle{plain}
\newtheorem{theorem}{\bf Theorem}
\newtheorem{lemma}[theorem]{\bf Lemma}
\newtheorem{corollary}[theorem]{\bf Corollary}
\newtheorem{proposition}[theorem]{\bf Proposition}
\newtheorem{definition}[theorem]{\bf Definition}
\newtheorem{condition}[theorem]{\bf Condition}
\newtheorem{remark}[theorem]{\it Remark}

\def\a{\alpha}  \def\cA{{\mathcal A}}     \def\bA{{\bf A}}  \def\mA{{\mathscr A}}
\def\b{\beta}   \def\cB{{\mathcal B}}     \def\bB{{\bf B}}  \def\mB{{\mathscr B}}
\def\g{\gamma}  \def\cC{{\mathcal C}}     \def\bC{{\bf C}}  \def\mC{{\mathscr C}}
\def\G{\Gamma}  \def\cD{{\mathcal D}}     \def\bD{{\bf D}}  \def\mD{{\mathscr D}}
\def\d{\delta}  \def\cE{{\mathcal E}}     \def\bE{{\bf E}}  \def\mE{{\mathscr E}}
\def\D{\Delta}  \def\cF{{\mathcal F}}     \def\bF{{\bf F}}  \def\mF{{\mathscr F}}
\def\c{\chi}    \def\cG{{\mathcal G}}     \def\bG{{\bf G}}  \def\mG{{\mathscr G}}
\def\z{\zeta}   \def\cH{{\mathcal H}}     \def\bH{{\bf H}}  \def\mH{{\mathscr H}}
\def\e{\eta}    \def\cI{{\mathcal I}}     \def\bI{{\bf I}}  \def\mI{{\mathscr I}}
\def\p{\psi}    \def\cJ{{\mathcal J}}     \def\bJ{{\bf J}}  \def\mJ{{\mathscr J}}
\def\vT{\Theta} \def\cK{{\mathcal K}}     \def\bK{{\bf K}}  \def\mK{{\mathscr K}}
\def\k{\kappa}  \def\cL{{\mathcal L}}     \def\bL{{\bf L}}  \def\mL{{\mathscr L}}
\def\l{\lambda} \def\cM{{\mathcal M}}     \def\bM{{\bf M}}  \def\mM{{\mathscr M}}
\def\L{\Lambda} \def\cN{{\mathcal N}}     \def\bN{{\bf N}}  \def\mN{{\mathscr N}}
\def\m{\mu}     \def\cO{{\mathcal O}}     \def\bO{{\bf O}}  \def\mO{{\mathscr O}}
\def\n{\nu}     \def\cP{{\mathcal P}}     \def\bP{{\bf P}}  \def\mP{{\mathscr P}}
\def\r{\rho}    \def\cQ{{\mathcal Q}}     \def\bQ{{\bf Q}}  \def\mQ{{\mathscr Q}}
\def\s{\sigma}  \def\cR{{\mathcal R}}     \def\bR{{\bf R}}  \def\mR{{\mathscr R}}
\def\S{\Sigma}  \def\cS{{\mathcal S}}     \def\bS{{\bf S}}  \def\mS{{\mathscr S}}
\def\t{\tau}    \def\cT{{\mathcal T}}     \def\bT{{\bf T}}  \def\mT{{\mathscr T}}
\def\f{\phi}    \def\cU{{\mathcal U}}     \def\bU{{\bf U}}  \def\mU{{\mathscr U}}
\def\F{\Phi}    \def\cV{{\mathcal V}}     \def\bV{{\bf V}}  \def\mV{{\mathscr V}}
\def\P{\Psi}    \def\cW{{\mathcal W}}     \def\bW{{\bf W}}  \def\mW{{\mathscr W}}
\def\o{\omega}  \def\cX{{\mathcal X}}     \def\bX{{\bf X}}  \def\mX{{\mathscr X}}
\def\x{\xi}     \def\cY{{\mathcal Y}}     \def\bY{{\bf Y}}  \def\mY{{\mathscr Y}}
\def\X{\Xi}     \def\cZ{{\mathcal Z}}     \def\bZ{{\bf Z}}  \def\mZ{{\mathscr Z}}
\def\O{\Omega}
\def\th{\theta}

\def\mx{{\mathscr x}}
\newcommand{\gA}{\mathfrak{A}}          \newcommand{\ga}{\mathfrak{a}}
\newcommand{\gB}{\mathfrak{B}}          \newcommand{\gb}{\mathfrak{b}}
\newcommand{\gC}{\mathfrak{C}}          \newcommand{\gc}{\mathfrak{c}}
\newcommand{\gD}{\mathfrak{D}}          \newcommand{\gd}{\mathfrak{d}}
\newcommand{\gE}{\mathfrak{E}}
\newcommand{\gF}{\mathfrak{F}}           \newcommand{\gf}{\mathfrak{f}}
\newcommand{\gG}{\mathfrak{G}}           
\newcommand{\gH}{\mathfrak{H}}           \newcommand{\gh}{\mathfrak{h}}
\newcommand{\gI}{\mathfrak{I}}           \newcommand{\gi}{\mathfrak{i}}
\newcommand{\gJ}{\mathfrak{J}}           \newcommand{\gj}{\mathfrak{j}}
\newcommand{\gK}{\mathfrak{K}}            \newcommand{\gk}{\mathfrak{k}}
\newcommand{\gL}{\mathfrak{L}}            \newcommand{\gl}{\mathfrak{l}}
\newcommand{\gM}{\mathfrak{M}}            \newcommand{\gm}{\mathfrak{m}}
\newcommand{\gN}{\mathfrak{N}}            \newcommand{\gn}{\mathfrak{n}}
\newcommand{\gO}{\mathfrak{O}}
\newcommand{\gP}{\mathfrak{P}}             \newcommand{\gp}{\mathfrak{p}}
\newcommand{\gQ}{\mathfrak{Q}}             \newcommand{\gq}{\mathfrak{q}}
\newcommand{\gR}{\mathfrak{R}}             \newcommand{\gr}{\mathfrak{r}}
\newcommand{\gS}{\mathfrak{S}}              \newcommand{\gs}{\mathfrak{s}}
\newcommand{\gT}{\mathfrak{T}}             \newcommand{\gt}{\mathfrak{t}}
\newcommand{\gU}{\mathfrak{U}}             \newcommand{\gu}{\mathfrak{u}}
\newcommand{\gV}{\mathfrak{V}}             \newcommand{\gv}{\mathfrak{v}}
\newcommand{\gW}{\mathfrak{W}}             \newcommand{\gw}{\mathfrak{w}}
\newcommand{\gX}{\mathfrak{X}}               \newcommand{\gx}{\mathfrak{x}}
\newcommand{\gY}{\mathfrak{Y}}              \newcommand{\gy}{\mathfrak{y}}
\newcommand{\gZ}{\mathfrak{Z}}             \newcommand{\gz}{\mathfrak{z}}

\def\ve{\varepsilon} \def\vt{\vartheta} \def\vp{\varphi}
\def\vk{\varkappa}
\def\vr{\varrho}

\def\be{{\bf e}} \def\bc{{\bf c}}
\def\bx{{\bf x}} \def\by{{\bf y}}
\def\bv{{\bf v}} \def\bu{{\bf u}}
 \def\bp{{\bf p}}
\def\mm{\mathrm m}
\def\mn{\mathrm n}

\def\ve{\varepsilon}   \def\vt{\vartheta}    \def\vp{\varphi}    \def\vk{\varkappa}

\def\Z{{\mathbb Z}}    \def\R{{\mathbb R}}   \def\C{{\mathbb C}}    \def\K{{\mathbb K}}
\def\T{{\mathbb T}}    \def\N{{\mathbb N}}   \def\dD{{\mathbb D}}


\def\la{\leftarrow}              \def\ra{\rightarrow}            \def\Ra{\Rightarrow}
\def\ua{\uparrow}                \def\da{\downarrow}
\def\lra{\leftrightarrow}        \def\Lra{\Leftrightarrow}


\def\lt{\biggl}                  \def\rt{\biggr}
\def\ol{\overline}               \def\wt{\widetilde}
\def\no{\noindent}


\let\ge\geqslant                 \let\le\leqslant
\def\lan{\langle}                \def\ran{\rangle}
\def\/{\over}                    \def\iy{\infty}
\def\sm{\setminus}               \def\es{\emptyset}
\def\ss{\subset}                 \def\ts{\times}
\def\pa{\partial}                \def\os{\oplus}
\def\om{\ominus}                 \def\ev{\equiv}
\def\iint{\int\!\!\!\int}        \def\iintt{\mathop{\int\!\!\int\!\!\dots\!\!\int}\limits}
\def\el2{\ell^{\,2}}             \def\1{1\!\!1}
\def\sh{\sharp}
\def\wh{\widehat}
\def\bs{\backslash}

\def\sh{\mathop{\mathrm{sh}}\nolimits}
\def\Area{\mathop{\mathrm{Area}}\nolimits}
\def\arg{\mathop{\mathrm{arg}}\nolimits}
\def\const{\mathop{\mathrm{const}}\nolimits}
\def\det{\mathop{\mathrm{det}}\nolimits}
\def\diag{\mathop{\mathrm{diag}}\nolimits}
\def\diam{\mathop{\mathrm{diam}}\nolimits}
\def\dim{\mathop{\mathrm{dim}}\nolimits}
\def\dist{\mathop{\mathrm{dist}}\nolimits}
\def\Im{\mathop{\mathrm{Im}}\nolimits}
\def\Iso{\mathop{\mathrm{Iso}}\nolimits}
\def\Ker{\mathop{\mathrm{Ker}}\nolimits}
\def\Lip{\mathop{\mathrm{Lip}}\nolimits}
\def\rank{\mathop{\mathrm{rank}}\limits}
\def\Ran{\mathop{\mathrm{Ran}}\nolimits}
\def\Re{\mathop{\mathrm{Re}}\nolimits}
\def\Res{\mathop{\mathrm{Res}}\nolimits}
\def\res{\mathop{\mathrm{res}}\limits}
\def\sign{\mathop{\mathrm{sign}}\nolimits}
\def\span{\mathop{\mathrm{span}}\nolimits}
\def\supp{\mathop{\mathrm{supp}}\nolimits}
\def\Tr{\mathop{\mathrm{Tr}}\nolimits}
\def\BBox{\hspace{1mm}\vrule height6pt width5.5pt depth0pt \hspace{6pt}}
\def\as{\text{as}}
\def\all{\text{all}}
\def\where{\text{where}}
\def\Dom{\mathop{\mathrm{Dom}}\nolimits}


\newcommand\nh[2]{\widehat{#1}\vphantom{#1}^{(#2)}}
\def\dia{\diamond}

\def\Oplus{\bigoplus\nolimits}



\def\qqq{\qquad}
\def\qq{\quad}
\let\ge\geqslant
\let\le\leqslant
\let\geq\geqslant
\let\leq\leqslant
\newcommand{\ca}{\begin{cases}}
\newcommand{\ac}{\end{cases}}
\newcommand{\ma}{\begin{pmatrix}}
\newcommand{\am}{\end{pmatrix}}
\renewcommand{\[}{\begin{equation}}
\renewcommand{\]}{\end{equation}}
\def\eq{\begin{equation}}
\def\qe{\end{equation}}
\def\[{\begin{equation}}
\def\bu{\bullet}
\def\tes{\textstyle}

\newcommand{\fr}{\frac}
\newcommand{\tf}{\tfrac}

\title[Inverse problem for the $L$-operator]
 {Inverse problem for the $L$-operator in the Lax Pair of
the Boussinesq equation on the circle}


\date{\today}
\author[Andrey Badanin]{Andrey Badanin}
\address{Saint-Petersburg
State University, Universitetskaya nab. 7/9, St. Petersburg, 199034
Russia, a.badanin@spbu.ru}
\author[Evgeny Korotyaev]{Evgeny L. Korotyaev}
\address{Saint-Petersburg
State University, Universitetskaya nab. 7/9, St. Petersburg, 199034
Russia, korotyaev@gmail.com}

\subjclass{47E05, 34L20, 34L40}
\keywords{inverse problem, eigenvalues,
3-rd order operator, Boussinesq equation}

\maketitle

\begin{abstract}
We consider a third-order non-self-adjoint operator, which is an
$L$-operator in the Lax pair for the Boussinesq equation on the
circle. We construct a mapping from the set of operator coefficients
to the set of spectral data, similar to the corresponding mapping
for the Hill operator constructed by E. Korotyaev. We prove that in
a neighborhood of zero our mapping is analytic and one-to-one.
\end{abstract}

\section{Introduction} Consider the third-order non-self-adjoint operator
$Hy=y'''+(py)'+py'+qy$
in $L^2(\R)$, with real periodic coefficients $p,q$.
We assume that
$$
\gu=(p,q)\in \gH:=\cH_1\os\cH,
$$
where $ \cH=\{f\in
L_\R^2(\T):\int_0^1f(x)dx=0\}$, $\cH_1=\{f:f,f'\in\cH\}$,
$\T=\R/\Z$.
The operator $H$ is an $L$-operator in the Lax pair for the Boussinesq equation
on the circle
$$
p_{tt}=-{1\/3}(p_{xxxx}+4(p^2)_{xx}), \qq q_x=p_t,\qq x\in\T.
$$
Kalantarov and Ladyzhenskaya \cite{KL77} proved that
solutions of the Boussinesq equation may have a blow up in a finite time.
McKean \cite{McK81} considered the operator $H$ with small coefficients
$p,q\in C^\iy(\T)$. McKean does not provide a solution
to the inverse problem in fixed smoothness classes, however, the paper \cite{McK81}
contains some ideas that we use in our solution of the inverse problem.

We study the inverse spectral problem in the class $\gH$ of the
coefficients. We construct a mapping from the space $\gH$ to the
space of spectral data and prove that this mapping is an analytic
bijection of some small ball in $\gH$ onto its image. We construct
our mapping as a composition of the Korotyaev's mapping for the Hill
operator from \cite{K99}, and the McKean's transformation from
\cite{McK81}. This transformation reduces the spectral problem for
the operator $H$ to the spectral problem for the Hill operator with
an energy-dependent potential.

\section{Mapping for the Hill operator}
There are two ways to solve the inverse problem for Hill operators.
First, we can apply the Marchenko-Ostrovsky results via the
conformal mapping \cite{MO75}. This method requires analysis of the
properties of the quasi-momentum. Second, we can construct the
Korotyaev map \cite{K99} using the lengths of the gaps. This mapping
is constructed only in spectral terms. The second method is more
convenient for the Boussinesq equation.

Let us recall how in \cite{K99} the mapping $v\to \{\text{\it
spectral data}\}$ is constructed for the Hill operator $-y''+vy,
v\in\cH$, acting on $L^2(\R)$. The spectrum consists of the
intervals $[\l_{n-1}^+,\l_n^-],n\in\N$, separated by the gaps
$(\l_n^-,\l_n^+)$, here $\l_0^+<\l_1^-\le \l_1^+<\l_2^-\le...$ are
the eigenvalues of the 2-periodic problem $y(0)=y(2),y'(0)=y'(2)$.
The eigenvalues $\gm_n,n\in\N$, of the Dirichlet problem
$y(0)=y(1)=0$ are simple and satisfy $\gm_n\in[\l_n^-,\l_n^+]$ for
all $n\in\N.$ The numbers $\gm_n,n\in\N$, are zeros of the entire
function $\vp(1,\cdot)$, where $\vp(x,\l)$ is the solution of the
equation $ -y''+vy=\l y$ under the initial conditions $\vp(0,\l
)=0,\vp'(0,\l )=1$. Introduce the mapping $\p:\cH\to
\ell^2\os\ell^2$ from \cite{K99} by
\[
\lb{defwtg}
\begin{aligned}
&\qqq\qqq\qqq\qqq\p(v)=(\p_n(v))_{n\in\N},\qqq\p_n=(\p_{cn},\p_{sn})\in\R^2,
\\
& \tes  \p_{cn}={1\/2}(\l_n^++\l_n^-)-\gm_n, \qq
\p_{sn}=\big|{1\/4}(\l_n^+-\l_n^-)^2-\p_{cn}^2\big|^{1\/2}\sign\gh_{sn},
\ \  \gh_{sn}=\log |\vp'(1,\gm_n)|,
\end{aligned}
\]
where the branch of the logarithm satisfies $\log 1=0$. Recall the
result of Korotyaev \cite{K98}, \cite{K99}.

\begin{theorem}
The mapping $\p$ is a real analytic isomorphism
between  $\cH$ and $\ell^2\os\ell^2$.
\end{theorem}

\section{Floquet multipliers and the 3-point problem}
Here we describe our spectral data.
Consider the equation
\[
\lb{1b}
y'''+(py)'+py'+qy=\l y,\qqq \gu=(p,q)\in \gH, \qq \l\in\C.
\]
Introduce the $3\ts 3$ {\it monodromy matrix} $
M(\l)=(\vp_j^{(k-1)}(1,\l))_{j,k=1}^3, $ where $\vp_1, \vp_2, \vp_3$
are fundamental solutions to Eq.~\er{1b}, satisfying the conditions
$\vp_j^{(k-1)}(0,\l)=\d_{jk},j,k=1,2,3 $.
The matrix function $M$ is entire and real for real $\l$.
The matrix $M$ has $3$ eigenvalues ({\it Floquet multiplier})
$\t_1, \t_2$, and $\t_3$, which satisfy the equality $\t_1\t_2\t_3=1. $
These three functions of the variable $\l$ form three different branches
of a function analytic on some 3-sheet Riemann surface $\cR$,
having only algebraic singularities in $\C$. {\it The branch points}
of the surface $\cR$ coincide with the zeros of the entire function
$
\r=(\t_1-\t_2)^2(\t_1-\t_3)^2(\t_2-\t_3)^2.
$

Introduce the operator
$H_{dir}y=y'''+(py)'+py'+qy$ on the interval $[0,2]$ under the
3-point Dirichlet conditions
$
y(0)=y(1)=y(2)=0.
$
The spectrum of $H_{dir}$ is pure discrete.

Introduce the domains $\cD_n, n\in\Z$, by
$$
\tes\cD_{n}=\{\l\in\C:|\l^{1\/3}-{2\pi
n\/\sqrt3}|<{2\/\sqrt3}\},\qq \cD_{-n}=\{\l\in\C:-\l\in\cD_n\},\qq
n\ge 0,
$$
here and below $\arg\l^{1\/3}\in(-{\pi\/3},{\pi\/3}]$.
Fix $\ve>0$ small enough and introduce the ball
$$
\tes \cB_\ve=\{\gu\in\gH:\int_0^1(|p'|^2+|q|^2)dx<\ve\}.
$$

\begin{theorem}
\lb{Th3pram}
Let $\gu\in\cB_\ve$. Then there are exactly two (counting with multiplicity)
zeros $r_n^\pm$ of the function $\r$ and exactly one simple eigenvalue
$\m_n$ of the operator  $H_{dir}$ in each domain
$\cD_n,n\in\Z_0$. There are no other eigenvalues of the operator
$H_{dir}$ and zeros of the function $\r$ in $\C$, with only exception
two zeros $r_0^\pm$ of the function $\r$ in $\cD_0$.
For all $n\in\Z_0$ the numbers $r_n^\pm$
and $\m_n$ are real, $\mu_{n}\in[r_{n}^-,r_{n}^+]$, and the following
identities hold true:
$$
r_n^\pm(\gu)=r_n^\pm(\gu^-)=-r_{-n}^\mp(\gu_*)=-r_{-n}^\mp(\gu_*^-),\qq
\m_{-n}(\gu)=-\m_n(\gu_*^-),
$$
where $ \gu_*=(p,-q), \gu^-(x)=\gu(1-x), \gu_*^-(x)=\gu_*(1-x). $

\end{theorem}

The eigenvalues $\m_n$ of the 3-point problem and the branch points
$r_n^\pm$ of the surface $\cR$ are our spectral data.
Let $y_n$ be the eigenfunctions of the 3-point problem corresponding
to the eigenvalues $\m_n$ and normalized by the condition $y_n'(0)=1$.

\section{The McKean transformation}
Our construction of the map $\gu\to\{spectral\ data\}$
is based on the corresponding Korotyaev map for the Hill operator
and uses the McKean transformation \cite{McK81} of our third-order
operator into the Hill operator with an energy-dependent potential.
We describe this transformation briefly.

Let $\gu\in\cB_\ve$. Then the multiplier $\t$, satisfying the
asymptotics $\t(\l)=e^{\l^{1\/3}}(1+o(1))$ as $\l\to+\iy$, is an
analytic function on
$\mD=\{\l\in\C:|\l|>1,\arg\l\in(-{3\pi\/4},{3\pi\/4})\}$ and $\t$
does not vanish in $\mD$. Moreover, $\t(\l)>0$ for all $\l>1$. Let
$\l\in\mD$. Then there exists the unique solution $\gf$ to
Eq.~\er{1b} such that $\gf(x+1,\l)=\t(\l)\gf(x,\l),\ \gf(0,\l)=1. $
Moreover, $\gf>0$ for all $\l>1$. Define the function $\gf^{1\/2}$
on $\R\ts\mD$ by the condition $\gf^{1\/2}>0$ at $\l>1$. If $y$ is a
solution to Eq.~\er{1b}, then the function $ u=\gf^{3\/2}({y\/\gf})'
$ satisfies the equation
\[
\lb{2oequd}
\tes -u''+V(x,E)u=E u,\qq V=E-2p-{3\/2}{\gf''\/\gf}
+{3\/4}({\gf'\/\gf})^2,
\]
where $ \tes E={3 \/4}\l^{2\/3}\in\O=\{E\in\C:|E|>1,\Re E>0\}. $
The energy-dependent 1-periodic potential $V(x,E),x\in\R$,
is a function analytic with respect to $E$ on the domain $\O$
and real at $E\in\R$.
Equations of the form \er{2oequd} were studied in our paper \cite{BK20}.
If  $\gu\in\cB_\ve$, then there are exactly two (counting with multiplicities)
eigenvalues $E_n^\pm$ of the 2-periodic problem $u(2)=u(0), u'(2)=u'(0)$
for Eq.~\er{2oequd} and exactly one simple eigenvalue $\gm_n$
of the Dirichlet problem $u(0)=u(1)=0$ in each domain
$$
\Omega_n
=\{E\in\C:|\sqrt E-\pi n|<1\},\qq n\in\N.
$$
There are no other eigenvalues in the domain $\O$.
All these eigenvalues are real and $\gm_n\in[E_n^-,E_n^+]$ for all $n\in\N.$
Introduce the solution $\vp(x,E)$ to Eq.~\er{2oequd},
satisfying the initial conditions $\vp(0,E)=0,\vp'(0,E)=1$.
Each of the functions $\vp(x,\cdot),x\in\R$, is analytic on $\Omega$.
The spectrum of the Dirichlet problem coincides with the set
of zeros of the function $\vp(1,\cdot)$ in $\Omega$.

\begin{theorem}
\lb{Th3}
Let $\gu\in\cB_\ve$, $n\in\N$. Then  $\vp(\cdot,\gm_n)$
is the eigenfunction of the Dirichlet problem, corresponding
to the eigenvalue $\gm_n$, and the following identities hold true:
\[
\lb{relr3p2o}
\begin{aligned}
\tes &\qqq\qqq\qqq E_n^\pm(\gu)={3\/4}(r_n^\pm(\gu))^{2\/3},\qq
\gm_n(\gu) ={3\/4}(-\m_{-n}(\gu_*))^{2\/3},
\\
\tes &\vp'(1,\gm_n(\gu))
=y_{-n}'(1,\gu_*)\t^{-{1\/2}}(-\m_{-n}(\gu_*),\gu),\qq
\t^{-{1\/2}}(-\m_{-n}(\gu_*),\gu)>0.
\end{aligned}
\]
\end{theorem}

Theorem~\ref{Th3} shows that the McKean transformation maps
our spectral data for the 3rd order operator to spectral data
for the Hill operator. Namely, for any $n\in\N$, the branch points
$r_n^\pm$ of the Riemann surface $\cR$ become the eigenvalues $E_n^\pm$
of the 2-periodic problem, and the eigenvalue $\m_n$ of the 3-point problem
becomes the eigenvalue $\gm_n$ of the Dirichlet problem.

\section{Mapping.}
McKeen \cite{McK81} proved the following result.

\no {\it
Let $(p_1,q_1),(p_2,q_2)\in\C^\iy(\T)\ts\C^\iy(\T)$,
be small enough and let the branch points of the surface $\cR$
and the eigenvalues of the 3-point problem on the surface
$\cR$ for $(p_1,q_1)$ and for $(p_2,q_2)$ coincide with each other.
Then $(p_1,q_1)=(p_2,q_2)$.
}

In this section we formulate Theorem~\ref{ThNablagsn} that strengthens
and refines McKean's result. We will construct our map
$\gu\to\{spectral\ data\}$ as a composition of
the Korotyaev map and the McKean transformation.
More precisely, substituting the identities \er{relr3p2o}
into \er{defwtg}, we obtain our map \er{gcn}.

Introduce the mapping $g:\cB_\ve\to \gh=\ell_1^2\os\ell_1^2$
by the identities $g(\gu)=(g_n(\gu))_{n\in\Z_0}$, where
$g_n=(g_{cn},g_{sn})\in \R^2$ and the functions $g_{cn}(\gu),g_{sn}(\gu),
\gu\in \cB_\ve$  have the form
\[
\lb{gcn}
\begin{aligned}
&\qqq\qqq\qqq\qqq\qqq g_n=(g_{cn}(\gu),g_{sn}(\gu)),\qq g_{-n}(\gu)=g_n(\gu_*^-),
\\
 & \tes
g_{cn}={3\/4}\big({1\/2}\big((r_n^+(\gu))^{2\/3}+(r_n^-(\gu))^{2\/3}\big)
-(-\m_{-n}(\gu_*))^{2\/3}\big),\qq
g_{sn}=\big|{1\/4}\g_n^2(\gu)-g_{cn}^2(\gu)\big|^{1\/2} \sign
h_{sn}(\gu),
\\
& \tes\qqq\qqq h_{sn}(\gu)=\log
|y_{-n}'(1,\gu_*)\t^{-{1\/2}}(-\m_{-n}(\gu_*),\gu)|, \qq
\g_n={3\/4}\big((r_n^+(\gu))^{2\/3}-(r_n^-(\gu))^{2\/3}\big).
\end{aligned}
\]
Here
$\ell_1^2=\ell_1^2(\Z_0)=\{(a_n)_{n\in\Z_0}:\sum_{n\in\Z_0}|na_n|^2<\iy\}$.

\begin{theorem}
\lb{ThNablagsn} The mapping $g:\cB_\ve\to \gh$ is a local analytic bijection
between $\cB_\ve$ and $g(\cB_\ve)$.
\end{theorem}

For the Boussinesq equation, this result means that solutions with sufficiently
small initial data maintain their smoothness class over time.

McKean \cite{McK81} found a one-band periodic solution
of the Boussinesq equation in terms of the Weierstrass function.
There are no results on the existence of N-band periodic solutions
for $N\ge 2$ in \cite{McK81}. Our Theorem~\ref{ThNablagsn}
implies the existence
of N-band solutions of the Boussinesq equations for any $N\ge 1$.

\begin{corollary}
\lb{Corfg}
The set of finite-band coefficients is dense in $\cB_\ve$.
\end{corollary}

\no\small {\bf Acknowledgments.}
The study was supported by a grant from
the Russian Science Foundation No. 23-21-00023,
https://rscf.ru/project/23-21-00023/


\begin{thebibliography}{9999}
\setlength{\itemsep}{-\parskip}
\footnotesize


\bibitem{BK20} A.Badanin, E.Korotyaev Hill's operators with
 the potentials analytically dependent on energy,
Journal of Differential Equations, 271 (2021), 638--664.


\bibitem{KL77} Kalantarov, V. K., Ladyzhenskaja, O. A.
Formation of collapses in quasilinear equations of parabolic and hyperbolic types.
(Russian) Boundary value problems of mathematical physics and related questions
in the theory of functions, 10.
Zap. Naucn. Sem. Leningrad. Otdel. Mat. Inst. Steklov.
(LOMI) 69 (1977), 77--102, 274.


\bibitem{K98}  E.Korotyaev,  Estimates of periodic potentials in terms
of gap lengths. Comm. Math. Phys. 197 (1998), no. 3, 521--526.

\bibitem{K99}  E.Korotyaev, Inverse Problem and
the trace formula for the Hill Operator, II, Mathematische
Zeitschrift 231(2) (1999), 345--368.


\bibitem{MO75}  Marchenko, V. A., Ostrovskii I. V.
Characteristics of the spectrum of the Hill operator. (Russian) Mat.
USSR Sb., 26 (1975), no.~4, 493--554.

\bibitem{McK81} H.McKean, Boussinesq's equation on the circle,
Com. Pure and Appl. Math. 34 (1981) 599--691.


\end{thebibliography}
\end{document}